\documentclass{article}

\usepackage{graphicx, amsmath, xcolor}
\usepackage{multicol, multirow, float, parskip, layout}
\usepackage{authblk}
\usepackage[a4paper, total={6in, 10in}]{geometry}
\usepackage{subcaption}
\usepackage{graphicx}

\DeclareCaptionFormat{custom}
{%
    \textbf{#1#2}\small #3
}
\title{An in-silico study of conventional and FLASH radiotherapy iso-effectiveness: Radiolytic oxygen depletion and its potential impact on tumor control probability}

\author[1, 2]{Isabel González-Crespo}
\author[3]{Faustino Gómez}
\author[1, 2]{Óscar López Pouso}
\author[1, 4, \thanks{{e-mail:juan.pardo.montero@sergas.es}}]{Juan Pardo-Montero}

\affil[1]{Group of Medical Physics and Biomathematics, Instituto de Investigación Sanitaria de Santiago (IDIS), Santiago de Compostela, Spain}
\affil[2]{Department of Applied Mathematics, Universidade de Santiago de Compostela, Santiago de Compostela, Spain}
\affil[3]{Department of Particle Physics, Universidade de Santiago de Compostela, Santiago de Compostela, Spain}
\affil[4]{Department of Medical Physics, Complexo Hospitalario Universitario de Santiago, Santiago de Compostela, Spain}

\begin{document}
\maketitle

\setlength{\parindent}{15pt}
\newpage	
	
\begin{abstract}

FLASH radiotherapy (FLASH-RT) has shown the potential to spare normal tissue while seemingly maintaining the effectiveness of conventional radiotherapy (CONV-RT). It has been suggested that the protective effect arises from the radiolytic oxygen depletion (ROD) caused by FLASH-RT, but it is not entirely clear why this protective effect is not observed in tumors. Iso-effectiveness has been experimentally observed in time-volume curves of preclinical tumors irradiated with FLASH and conventional radiotherapy, but it may not translate to clinical trials, where tumor control probability (TCP) is typically the investigated endpoint.\\

In this work, we used mathematical models to investigate the iso-effectiveness of FLASH-RT/CONV-RT on tumors, focusing on the role of ROD. We used a spatiotemporal reaction-diffusion model, including ROD, to simulate tumor oxygenation. From those oxygen distributions we obtained surviving fractions (SFs), using the linear-quadratic model with oxygen enhancement ratios (OER). We then used the calculated SFs to describe the evolution of preclinical tumor volumes through a mathematical model of tumor response. We also calculated TCPs using the Poisson-LQ approach. \\

Our study suggests that ROD causes differences in SF between FLASH-RT and CONV-RT, especially in low $\alpha$/$\beta$ and poorly oxygenated cells. These changes do not lead to significant differences in the evolution of preclinical tumors. However, when extrapolating this effect to TCP curves, we observed important differences between both techniques (TCP is lower in FLASH-RT). Nonetheless, it cannot be discarded that other effects not modeled in this work could contribute to tumor control and maintain the iso-effectiveness of FLASH-RT.

\end{abstract}

\newpage

\section{Introduction}
\label{section_intro}

In FLASH radiotherapy (FLASH-RT) the radiation dose is delivered at ultra-high dose rates (UHDR) exceeding 40~Gy/s, in contrast to the 0.05-0.4~Gy/s used in conventional radiotherapy \cite{durante2018, favaudon2022}. FLASH-RT has generated significant interest in recent years, as \textit{in vivo} experiments have shown its potential to spare non-tumor tissue while maintaining the effectiveness of conventional radiotherapy (CONV-RT) on tumors, known as the FLASH effect~\cite{favaudon2014, montay2017, montay2018, vozenin2019, montay2019, levy2020, diffenderfer2020, liljedahl2022, gao2022}.

Many studies have suggested that the protective effect arises from the radiolytic oxygen depletion (ROD) process caused by FLASH irradiation~\cite{favaudon2022, pratx2019a, pratx2019b, petersson2020, cao2021, el2022, ha2022,jansen2022}, that results in increased cell radioresistance. However, other research works propose that oxygen depletion alone is insufficient to fully explain the sparing effect~\cite{spitz2019, jin2020, abolfath2022}. Instead, these studies suggest that additional physicochemical factors may contribute to the phenomenon, such as variations in reactive species production between tumor and non-tumor cells, the recombination of free radicals, or radiation-induced immune effects.

Understanding why the protective effect is not observed in tumors remains an ongoing area of research, with potentially important clinical implications: it is important to guarantee that clinical trials of FLASH-RT will not lead to poor tumor control probabilities (TCP), and if necessary to boost the radiation dose to compensate for such effects.   

In this regard, a recent theoretical study by Liew~\textit{et al.}~\cite{liew2023} has argued that FLASH-RT might not preserve tumor control probability. In this independent study, we further investigate this possibility, theoretically analyzing the effect of the oxygen depletion associated with FLASH-RT on the surviving fraction of cells, how this can affect the growth curves of preclinical tumors, and finally the TCP-dose curves of tumors according to their oxygenation status and radiosensitivity. The mathematical modeling is based on the differential equations governing the oxygen distribution in tissues, the linear-quadratic (LQ) model with oxygen enhancement ratios (OERs) to account for the oxygen effect on irradiated cells, a simple model of tumor growth after irradiation, and the standard LQ-Poisson TCP formulation to calculate the tumor control probability of different treatments. We first present the models in detail, and then we progressively analyze experimental data of ROD in solutions, ROD \textit{in vivo}, surviving fractions \textit{in vitro}, and tumor growth in preclinical irradiated tumors. We finally extrapolate the results obtained to the analysis of TCP-dose curves in a clinical setting, and the potential effect of ROD on the clinical iso-effectiveness of FLASH-RT.

\section{Materials and Methods}
\label{section_materials}

\subsection{Oxygen depletion in FLASH-RT and CONV-RT}
\label{section_materials_pO2}

Many experimental studies suggest that the amount of depleted oxygen by FLASH-RT depends on the initial oxygen partial pressure~\cite{el2022, jansen2022, van2022}. These works indicate that there is a pronounced curvature in the depletion pattern as the oxygen partial pressure approaches zero. Furthermore, Cao \textit{et al.}~\cite{cao2021} conducted an \textit{in vivo} experiment and found that oxygen depletion does not depend on the radiation dose rate at UHDR, within the 50--300~Gy/s interval, but it does on the radiation dose. As it was done in previous works~\cite{jansen2022, taylor2022}, we used a Michaelis-Menten kinetics to describe oxygen depletion \textit{in vitro} and in aqueous solutions (with homogeneous oxygen initial conditions):
\begin{equation}
	\frac{dp(t)}{dt}= \frac{D}{T}\frac{G_0\ p(t)}{k_\mathrm{ROD}+p(t)},
	\label{eq_michaelis}
\end{equation}
being $p(t)$ is the time-dependent oxygen partial pressure, $D$ is the radiation dose, $T$ is the total time of irradiation, $G_0$ is the radiolytic consumption rate and $k_\mathrm{ROD}$ is the oxygen pressure for half-maximum oxygen depletion rate. This expression allowed us to fit experimental data of oxygen variations and to obtain optimal values for the parameters $G_0$ and $k_\mathrm{ROD}$.

Following the work of Taylor \textit{et al}.~\cite{taylor2022}, we used a spatiotemporal reaction-diffusion equation to model the dynamics of oxygen in tissues with heterogeneous oxygenation (as preclinical tumors):
\begin{equation}
	\frac{\partial p(\mathbf{x},t)}{\partial t} = D_\mathrm{O_2} \Delta p(\mathbf{x},t) - g_\mathrm{max}\frac{p(\mathbf{x},t)}{k+p(\mathbf{x},t)}-\frac{D}{T}G_0\frac{p(\mathbf{x},t)}{k_\mathrm{ROD}+p(\mathbf{x},t)},
	\label{eq_oxyFlash}
\end{equation}
where $t$ and $\mathbf{x}$ are the time and spatial coordinates, $D_\mathrm{O_2}$ is the oxygen diffusion coefficient, $\Delta$ is the Laplacian operator with respect to $\mathbf{x}$, $g_\mathrm{max}$ is the maximum metabolic consumption rate, $k$ is the oxygen pressure for half-maximum consumption rate and the last term is the Michaelis-Menten term given by Equation~(\ref{eq_michaelis}).

We used the Partial Differential Equation Toolbox of MATLAB~\cite{partialToolbox} to solve Equation~(\ref{eq_oxyFlash}) by using a Finite Element Method (FEM) in a two-dimensional space domain, building on a solver that was developed in a previous work~\cite{rodriguez2019} to simulate the tumor vasculature and oxygenation in a two-dimensional domain. The two-dimensional problem was shown to be a good surrogate of the three-dimensional problem~\cite{espinoza2013}. Blood capillaries were randomly placed in the domain (avoiding overlapping and maintaining inter-distance restrictions) according to a given vascular fraction, $\textit{vf}$, defined as the ratio of capillary area to total area. Capillaries were assumed to be cylindrical and to follow a lognormal distribution of radii, based on the experimental characterization of the vascular network of tumors reported in~\cite{konerding2013}. In this work, we employed a distribution of capillaries obtained from the analysis of experimental data reported in~\cite{konerding2013} for a model of colon carcinoma, characterized by the following parameters: minimum diameter, 0.9 $\mu$m; maximum diameter, 161.9 $\mu$m; mean diameter, 65.4 $\mu$m; standard deviation, 50 $\mu$m (the mean and standard deviation both refer to the lognormal distribution).

On the outer boundary of the domain, zero-flux Neumann conditions were imposed on the oxygen partial pressure, while Dirichlet conditions with $p$=40~mmHg were set on the capillaries. As argued by Taylor~\textit{et al}.~\cite{taylor2022}, it was assumed that capillary oxygen partial pressure remains unaltered by FLASH-RT.

The oxygen partial pressure remains near-constant during CONV-RT \textit{in vivo}~\cite{cao2021, el2022, van2022}: as radiation is delivered in low dose rates, it leads to a slow radiolytic oxygen depletion, which is immediately compensated with the diffusion of oxygen from the vascular system. Consequently, the stationary solution of Equation~(\ref{eq_oxyFlash}), excluding the radiolytic consumption term, gives the distribution of oxygen partial pressure during CONV-RT and serves as an initial condition for solving the FLASH-RT evolutionary problem:
\begin{equation}
    D_\mathrm{O_2}\Delta p(\mathbf{x}) = g_\mathrm{max}\frac{p(\mathbf{x})}{k+p(\mathbf{x})}.
    \label{eq_oxySteady}
\end{equation}
 
In Figure \ref{fig_s1} we illustrate the solutions to Equation (\ref{eq_oxySteady}) on a pixel of 1$\times$1 mm$^2$~\cite{rodriguez2019} for different vascular fractions, leading to poorly and well oxygenated tumors. In Figure \ref{fig_s2} we illustrate the solution of Equation (\ref{eq_oxyFlash}) with a dose of 30 Gy (100 Gy s$^{-1}$) \textit{in vivo} (the capillaries act as a source and the initial oxygenation is recovered) and \textit{in vitro} (no recovery).

\subsection{Surviving fractions in FLASH-RT and CONV-RT}
\label{section_materials_SF}

We used the LQ model~\cite{fowler1989} with the OER modification~\cite{wouters1997} to obtain surviving fractions (SFs):
\begin{equation}
    \mathit{SF(p,D)} = \exp (-\alpha(p)D-\beta(p)D^2).
    \label{eq_LQ}
\end{equation}
The radiation damage parameters, $\alpha$ and $\beta$, depend on the oxygen partial pressure:
\begin{eqnarray}
	&\displaystyle \mathit{\alpha(p)} = \frac{\alpha_{\mathrm{ox}}}{{\mathit{OER}_\alpha}}\frac{{\mathit{OER}_\alpha} p  + k_\mathrm{m}}{p  + k_\mathrm{m}}, \label{eq_OERa}\\
	&\displaystyle \mathit{\beta(p)} = \frac{\beta_{\mathrm{ox}}}{{\mathit{OER}_\beta}^2}\frac{({\mathit{OER}_\beta} p  + k_\mathrm{m})^2}{(p  + k_\mathrm{m})^2},
	\label{eq_OERb}
\end{eqnarray}
being $\alpha_\mathrm{ox}$ and $\beta_\mathrm{ox}$ the $\alpha$ and $\beta$ parameters under fully aerobic conditions; ${\mathit{OER}_\alpha}$ and  ${\mathit{OER}_\beta}$ the maximum oxygen enhancement ratios; and $k_\mathrm{m}$ the oxygen partial pressure at which OERs achieve the half-maximum value.

The surviving fraction after a CONV-RT and FLASH-RT irradiation ($\mathit{SF}_\mathrm{C}$ and $\mathit{SF}_\mathrm{F}$, respectively) are calculated as follows~\cite{taylor2022}:
\begin{eqnarray}
	&\displaystyle \mathit{SF}_\mathrm{C} = \int_{0}^{\infty}\mathit{SF}(p,D) f(p)dp,
	\label{eq_SF_C}\\
	&\displaystyle \mathit{SF}_\mathrm{F} = \frac{1}{T} \int_{0}^{T}\left(\int_{0}^{\infty}\mathit{SF}(p,D) f(p,t) dp\right) dt,
	\label{eq_SF_F}
\end{eqnarray}
where $f(p,t)$ is the time-dependent distribution of oxygen partial pressure, obtained from solving Equation~(\ref{eq_oxyFlash}), and $f(p)$ is the analog for the steady-state case, given by Equation~(\ref{eq_oxySteady}).

\subsection{Model of tumor volume evolution}
\label{section_materials_model}

We used a simple compartmental mathematical model to describe the evolution of tumor volumes after FLASH-RT/CONV-RT treatments. We considered two populations of tumor cells, viable and radiation-doomed, that occupy volumes $C$, and $C_\mathrm{d}$, respectively. The cells undergo logistic proliferation with growth rate $\lambda$ and carrying capacity $K$. Moreover, doomed cells are eliminated with a depletion rate $\gamma$, which is constrained to be higher than $\lambda$ (we can define a positive depletion rate $\phi=\gamma-\lambda$), therefore resulting in a progressive elimination of damaged cells:
\begin{eqnarray}
	&\displaystyle \frac{dC(t)}{dt} = \lambda\left(1-\frac{C(t)+C_\mathrm{d}(t)}{K}\right)C(t),
	\label{eq_C}\\
	&\displaystyle \frac{dC_\mathrm{d}(t)}{dt} = - \lambda\frac{C(t)+C_\mathrm{d}(t)}{K}C_\mathrm{d}(t) -\phi C_\mathrm{d}(t).
	\label{eq_Cd}
\end{eqnarray}
The total volume, $V$, is defined as the sum of the two compartments, $V(t)=C(t)+C_\mathrm{d}(t)$.

We model the radiation effect as an \textit{impulse} in the above equations (impulsive differential equations). Let $\{t_i, i=1\dots n\}$ be the set of the $n$ days in which a radiation treatment is delivered. Equations (\ref{eq_C}) and (\ref{eq_Cd}) are defined for $t\neq t_i$, while at the irradiation times, the following impulsive terms are added to the model:
\begin{eqnarray}
	&\displaystyle \Delta C(t_i) = (\mathit{SF}-1)C(t_i),
	\label{eq_sf_C}\\
	&\displaystyle \Delta C_\mathrm{d}(t_i) = (1-\mathit{SF})C(t_i),
	\label{eq_sf_Cd}
\end{eqnarray}
where $\mathit{SF}$ is the corresponding surviving fraction for CONV-RT or FLASH-RT, obtained with Equation~(\ref{eq_SF_C})~or~(\ref{eq_SF_F}), respectively.

\subsection{Tumor control probability versus radiation dose}
\label{section_materials_TCP}

We used the Poisson-LQ approach~\cite{webb1993} to calculate tumor control probabilities:
\begin{equation}
	\mathit{TCP}= \exp(-N_0\mathit{SF}),
	\label{eq_tcp}
\end{equation}
where $N_0$ is the number of clonogenic cells within the tumor at the irradiation time and $\mathit{SF}$ is the surviving fraction after a treatment, obtained from Equations (\ref{eq_SF_C}) and (\ref{eq_SF_F}).

In order to simulate heterogeneous populations of tumors, we generated sets of 1000 \textit{tumors} by randomly assigning different response parameters to each of them. In particular, each tumor was characterized by a different $\alpha_\mathrm{ox}$ and $N_0$ (sampled from a normal distribution with standard deviation equal to 20\% of the central value), and a different oxygen distribution and radiolytic consumption rate\footnote{Oxygen distribution and radiolytic consumption rate for the 1000 tumors were sampled from a list of 100 distributions generated by solving Equation (\ref{eq_oxyFlash}). This step was limited to 100 simulations due to the computational cost of solving Equation (\ref{eq_oxyFlash}). For each simulation a different vascular fraction and different distribution of capillaries was used, in order to have tumors with different levels of oxygenation, ranging from highly hypoxic to well oxygenated. Associated to each solution there was also a different value of $G_0$ sampled from a normal distribution with 0.25~mmHg/Gy mean and standard deviation equal to 20\% of the central value.}.

Tumor control probabilities were calculated for several doses (single fraction treatments) in order to obtain the TCP-dose curves. Such curves were then characterized by calculating $D_{50}$ and $D_{90}$, the doses achieving $\mathit{TCP}$=0.5 and 0.9, respectively, by interpolating the TCP-dose curve.

\subsection{Experimental data collection, model fitting, and statistical analysis}
\label{section_materials_data}

We used the G3Data Graph Analyzer to collect relevant data from experimental research works for model fitting. In particular:

\begin{itemize}
    \item ROD data: references~\cite{el2022, jansen2022, van2022} reported measurements of the amount of depleted oxygen by FLASH-RT in model solutions with different dose rates and baseline oxygen levels. Moreover, Van Slyke \textit{et al}.~\cite{van2022} reported measurements \textit{in vivo} of oxygen depletion during FLASH-RT, delivering $30$~Gy with a dose rate of $100$~Gy/s on preclinical tumors with different oxygen distributions.
    
    \item Tumor volume evolution data: the study conducted by Diffenderfer~\textit{et~al.}~\cite{diffenderfer2020} involved 64 mice separated in 5 groups: control (5 mice), 12~Gy CONV-RT (15 mice), 12~Gy FLASH-RT (15 mice), 18~Gy CONV-RT (15 mice) and 18~Gy FLASH-RT (14 mice). The dose rate used in those FLASH-RT experiments was 63~Gy/s. Zhu~\textit{et~al.}~\cite{zhu2023a} conducted two independent studies, each of them with a different tumor line, involving 24 mice in each experiment separated in 3 groups: control (8 mice), 9.5~Gy CONV-RT (8 mice) and 9.5~Gy FLASH-RT (8 mice). In this case, the dose rate was 125~Gy/s.
\end{itemize}

We used the nonlinear optimization tool \texttt{fmincon} from the MATLAB Optimization Toolbox to fit Equation~(\ref{eq_michaelis}) to the reported datasets of oxygen consumption during FLASH-RT. We also implemented a \textit{Simulated Annealing} (SA) algorithm~\cite{kirkpatrick1983} to fit the model given by Equations~(\ref{eq_C})--(\ref{eq_sf_Cd}) to the datasets of volume dynamics. In both cases, our objective function was the sum of squared differences between the experimental data and the predicted value obtained with the models.

%
%

We used ANCOVA to study whether the differences in the volume dynamics curves between the control, CONV-RT and FLASH-RT groups were statistically significant~\cite{reichardt2019}. As ANCOVA uses linear models and tumor growth presents an exponential-like shape, we worked with the logarithm of tumor volumes, as it was done in~\cite{favaudon2014}. For this study, we used a separate lines model with the \texttt{aoctool} function from the MATLAB Statistics and Machine Learning Toolbox. Moreover, we used the \texttt{multcompare} function to perform multiple comparison tests and obtain p-values. We considered that two curves were significantly different if $\textrm{p-value}<0.05$.

\subsection{Parameter values}

Some of the model parameters presented before were fitted to experimental data and others were varied to analyze their effect on the results. However, some parameters were set to fixed values typically used in the radiobiological literature, reported in Table \ref{table_0}.

\begin{table}[htb]
    \centering
    \begin{tabular}{ccc}
        \cline{1-3}
	Parameter              & Value            & Equations\\\hline
       $g_\mathrm{max}$        & 15 mmHg s$^{-1}$ & (\ref{eq_oxyFlash}, \ref{eq_oxySteady}) \\
       $k$                     & 2.5 mmHg         & (\ref{eq_oxyFlash}, \ref{eq_oxySteady}) \\
       $D_\mathrm{O_2}$        & 2$\times$10$^{-9}$ m$^2$ s$^{-1}$ & (\ref{eq_oxyFlash}, \ref{eq_oxySteady}) \\
       ${\mathit{OER}_\alpha}$ & 2.5              & (\ref{eq_OERa}, \ref{eq_OERb}) \\
       ${\mathit{OER}_\beta}$  & 2.5              & (\ref{eq_OERa}, \ref{eq_OERb})  \\
       $k_\mathrm{m}$          & 3.28 mmHg        & (\ref{eq_OERa}, \ref{eq_OERb})  \\\hline
    \end{tabular}
    \caption{List of parameter values used in this work.}
    \label{table_0}
\end{table}

\section{Results and Discussion}
\label{section_results}


\subsection{Simulation of oxygen depletion in FLASH-RT}
\label{section_results_pO2}

\begin{figure}[htb]
	\centering
	\includegraphics[width=\columnwidth]{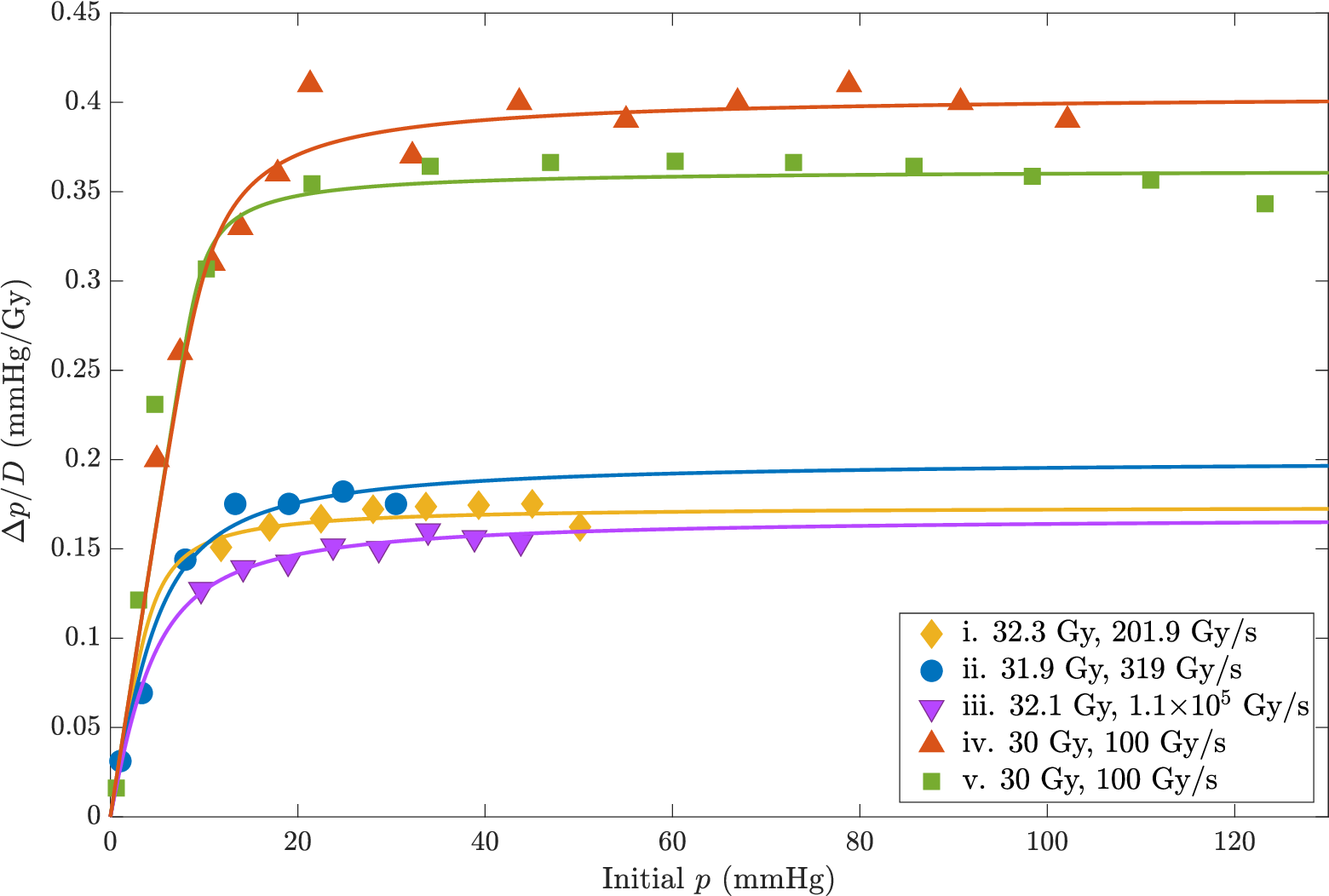}
	\caption{Oxygen depletion curves in different solutions, obtained by fitting Equation~(\ref{eq_michaelis}) to the measurements reported by Jansen~\textit{et~al.}~\cite{jansen2022} (i, ii and iii), Van Slyke~\textit{et~al.}~\cite{van2022} (iv) and El Khatib~\textit{et~al.}~\cite{el2022} (v). It is represented the total amount of depleted oxygen during FLASH-RT, $\Delta p$, divided by the delivered radiation dose against the initial oxygen partial pressure.}
	\label{fig_1}
\end{figure}

We fitted Equation~(\ref{eq_michaelis}) to the oxygen depletion measurements in solutions reported in references \cite{el2022, jansen2022, van2022} to obtain the best-fitting values of $G_0$ and $k_\mathrm{ROD}$. For that purpose, we used the fitting tools described in Section~\ref{section_materials_data}. In Figure~\ref{fig_1} and Table~\ref{table_1}, we show the fitted curves and the best-fitting parameter values, respectively.

\begin{table}[htb]
		\centering
		\begin{tabular}{ccc}
			\cline{1-3}
		                            &$G_0$ (mmHg/Gy) &$k_\mathrm{ROD}$ (mmHg)\\\hline
			\multicolumn{1}{l}{curve i}   & 0.18           & 1.51              \\ 
			\multicolumn{1}{l}{curve ii}  & 0.20           & 2.42              \\ 
			\multicolumn{1}{l}{curve iii} & 0.17           & 2.27              \\ 
			\multicolumn{1}{l}{curve iv}  & 0.36           & 0.65              \\ 
			\multicolumn{1}{l}{curve v}   & 0.41           & 1.30              \\ \hline
		\end{tabular}
	\caption{Best-fitting values of the parameters $G_0$ and $k_\mathrm{ROD}$ obtained by fitting Equation~(\ref{eq_michaelis}) to the oxygen depletion measurements reported in references~\cite{el2022, jansen2022, van2022}. The fits are shown in Figure~\ref{fig_1}.}
	\label{table_1}
\end{table}

Subsequently, we used Equation~(\ref{eq_oxyFlash}) to qualitatively fit the \textit{in vivo} experimental data presented by Van Slyke \textit{et al}.~\cite{van2022}. As preclinical tumors exhibit non-homogeneous oxygen distributions, we first followed the methodology described in Section~\ref{section_materials_pO2} to obtain a representative sample, $\Omega$, of $100$ tumors with mean oxygen levels in the range of $\simeq$[5, 30]~mmHg, similar to those reported in the literature~\cite{van2022}. Each oxygen distribution in the $\Omega$ sample was obtained from a random vascular fraction, sampled from a normal distribution with $\mu$=0.1, $\sigma$=0.04, and a 0.04 cutoff. The characteristics of the capillary geometry were those presented in Section \ref{section_materials_pO2}. We also assigned a different value of $G_0$ to each oxygen distribution in the $\Omega$ sample in order to introduce variability in the depletion rates. $G_0$ was sampled from a normal Gaussian distribution with a mean of 0.25~mmHg/Gy, near to the mean value of the $G_0$ values reported in Table~\ref{table_1}, and standard deviation equal to 25~\% of the mean. We set $k_\mathrm{ROD}=1$~mmHg, a value qualitatively similar to those reported in Table~\ref{table_1} and previous modeling studies~\cite{taylor2022}.

\begin{figure}[htb]
	\centering
	\includegraphics[width=\columnwidth]{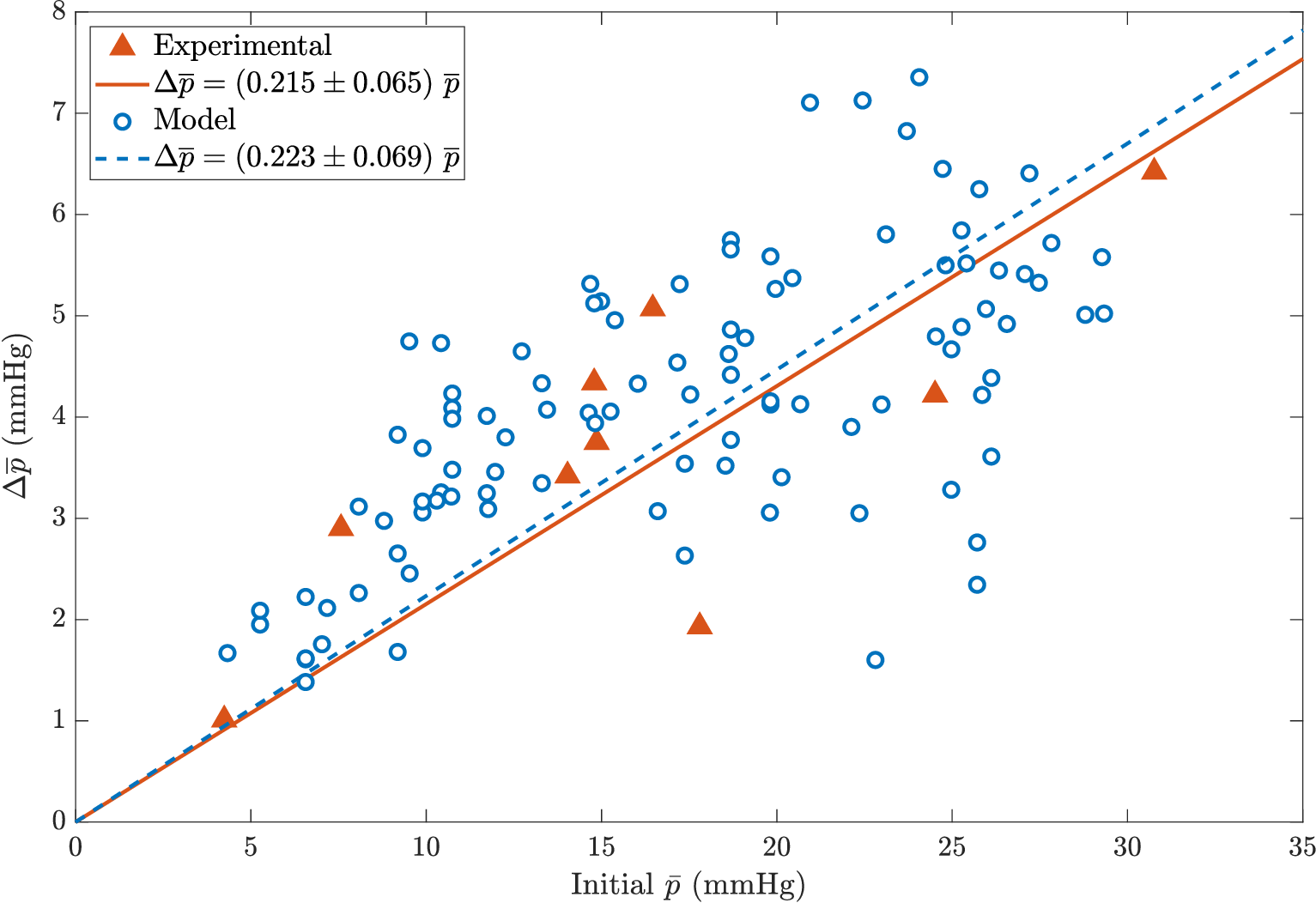}
	\caption{Amount of depleted oxygen during FLASH-RT versus the initial oxygen partial pressure in preclinical tumors. The triangles represent \textit{in vivo} data reported in~\cite{van2022}, and the circles represent the simulated data for 100 tumors with different oxygenations distributed within the experimental range, obtained by solving Equation~(\ref{eq_oxyFlash}). The linear fitting of each data set is presented as a solid line and a dashed line, respectively. The regression slopes of the linear fit of experimental and simulated values are also reported.}
	\label{fig_2}
\end{figure}

In Figure~\ref{fig_2} we present the simulated and experimental data. There is a qualitatively good agreement between experiment and model, including the slope of the curve and the dispersion of data.

\subsection{Comparison of surviving fractions in CONV-RT and FLASH-RT}
\label{section_results_SF}

Based on the characterization of oxygen depletion in FLASH-RT presented on the previous section, we investigated the potential effect on the surviving fraction of irradiated cells in \textit{in vitro} experiments. We set $G_0 = 0.25$~mmHg/Gy and $k_\text{ROD}=1$~mmHg and obtained the amount of depleted oxygen during FLASH-RT for different radiation doses and baseline oxygen levels in the [1, 40]~mmHg interval by using Equation~(\ref{eq_michaelis}). We limited the study to up to 40~mmHg,  as it was previously reported (both from experimental and modeling studies~\cite{adrian2020, zhu2021}) that for oxygen pressures above 30--40~mmHg the FLASH effect was not observed. Then, we calculated the associated SF with both CONV-RT and FLASH-RT using Equations~(\ref{eq_SF_C}) and (\ref{eq_SF_F}), respectively. We performed this study for different values of $\alpha_\mathrm{ox}$/$\beta_\mathrm{ox}$, 3, 10, 20, and $\infty$ Gy, setting ($\alpha_\mathrm{ox}$=0.1565 Gy$^{-1}$, $\beta_\mathrm{ox}$=0.0522 Gy$^{-2}$), ($\alpha_\mathrm{ox}$=0.4 Gy$^{-1}$, $\beta_\mathrm{ox}$=0.04 Gy$^{-2}$), ($\alpha_\mathrm{ox}$=0.6 Gy$^{-1}$, $\beta_\mathrm{ox}$=0.03 Gy$^{-2}$), and ($\alpha_\mathrm{ox}$=1.2 Gy$^{-1}$, $\beta_\mathrm{ox}$=0 Gy$^{-2}$)\footnote{These values of $\alpha_\mathrm{ox}$ and $\beta_\mathrm{ox}$ were set to be iso-effective at 30 Gy.}, respectively.

\begin{figure}[htb]
	\centering
	\includegraphics[width=\columnwidth]{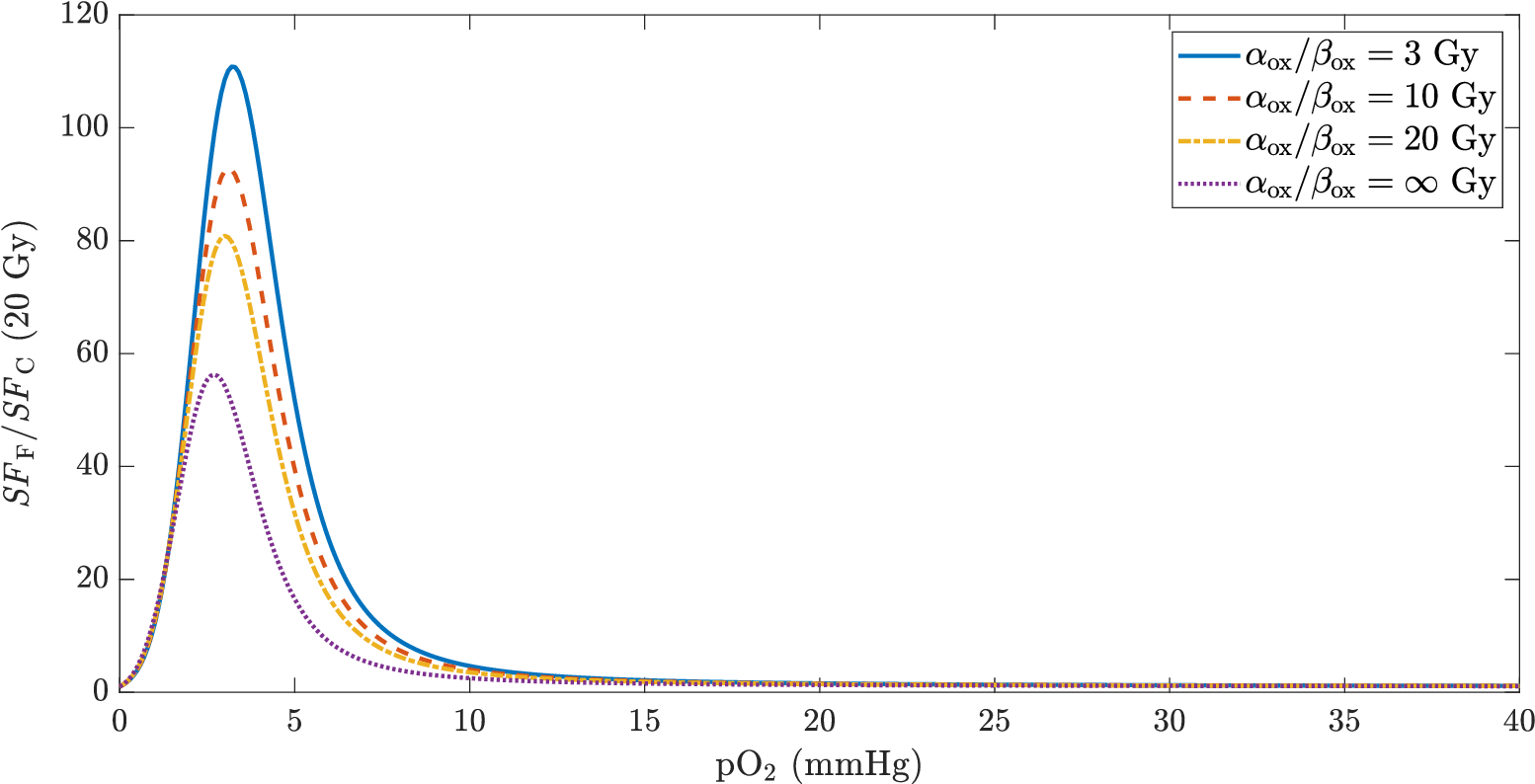}
	\caption{Ratio of surviving fractions for FLASH-RT ($\textit{SF}_\text{F}$) and CONV-RT ($\textit{SF}_\text{C}$) for a dose of 20~Gy versus the oxygenation of the cells.}
	\label{fig_SF}
\end{figure}

The surviving fraction versus dose curves are shown in Figure \ref{fig_s3}. The differences in SF between FLASH-RT and CONV-RT match qualitatively well with previously reported experimental studies, which showed higher SFs in FLASH-RT and revealed the influence of oxygen on these differences~\cite{adrian2020}. In Figure~\ref{fig_SF}, we illustrate the difference in SF between CONV-RT and FLASH-RT versus the oxygenation for a radiation dose of 20~Gy. It is observed that differences tend to zero in the limits $p\to \infty$ and $p\to 0$, the reason being that at large oxygen concentrations, the effect of ROD on the OERs becomes insignificant and that ROD is limited at very low oxygen concentrations. The maximum differences are observed approximately in the range [2, 5] mmHg. The location of this range depends on both the modeling of ROD and the modeling of the OER, particularly in the parameter $k$, which was set to 3.28 mmHg in this work.

\subsection{Fitting of the response model to preclinical tumor volume curves}
\label{section_results_preclinical}

We used the SA algorithm, as mentioned in Section~\ref{section_materials_data}, to fit the tumor response model given by Equations~(\ref{eq_C})--(\ref{eq_sf_Cd}) to datasets of tumor volume dynamics after CONV-RT and FLASH-RT reported in preclinical studies~\cite{diffenderfer2020, zhu2023a}. We included $\alpha_\mathrm{ox}$, $\lambda$, $K$, $\phi$ and the initial volume, $V_0=C(0)$, as fitting parameters (being $C_{\rm{d}}(0)=0$, and $t$=0 marking the irradiation time), and we fixed $\alpha_\mathrm{ox}/\beta_\mathrm{ox}$=10~Gy. To take into account the variability in both the oxygen distribution and the radiolytic consumption rate, we performed an optimization process for each oxygen distribution in the $\Omega$ sample with the $G_0$ parameter assigned for the heterogeneous study described in Section~\ref{section_results_pO2}. We then selected the distribution and optimal parameters that minimize the cost function, defined in Section~\ref{section_materials_data}\footnote{However, differences in the value of the cost function of each of the 100 optimizations were minimal, and the only difference in the best-fitting parameters was an increasing value of $\alpha_\mathrm{ox}$ with decreasing oxygenation.}.

\begin{figure}[htb]
	\centering
	\includegraphics[width=\textwidth]{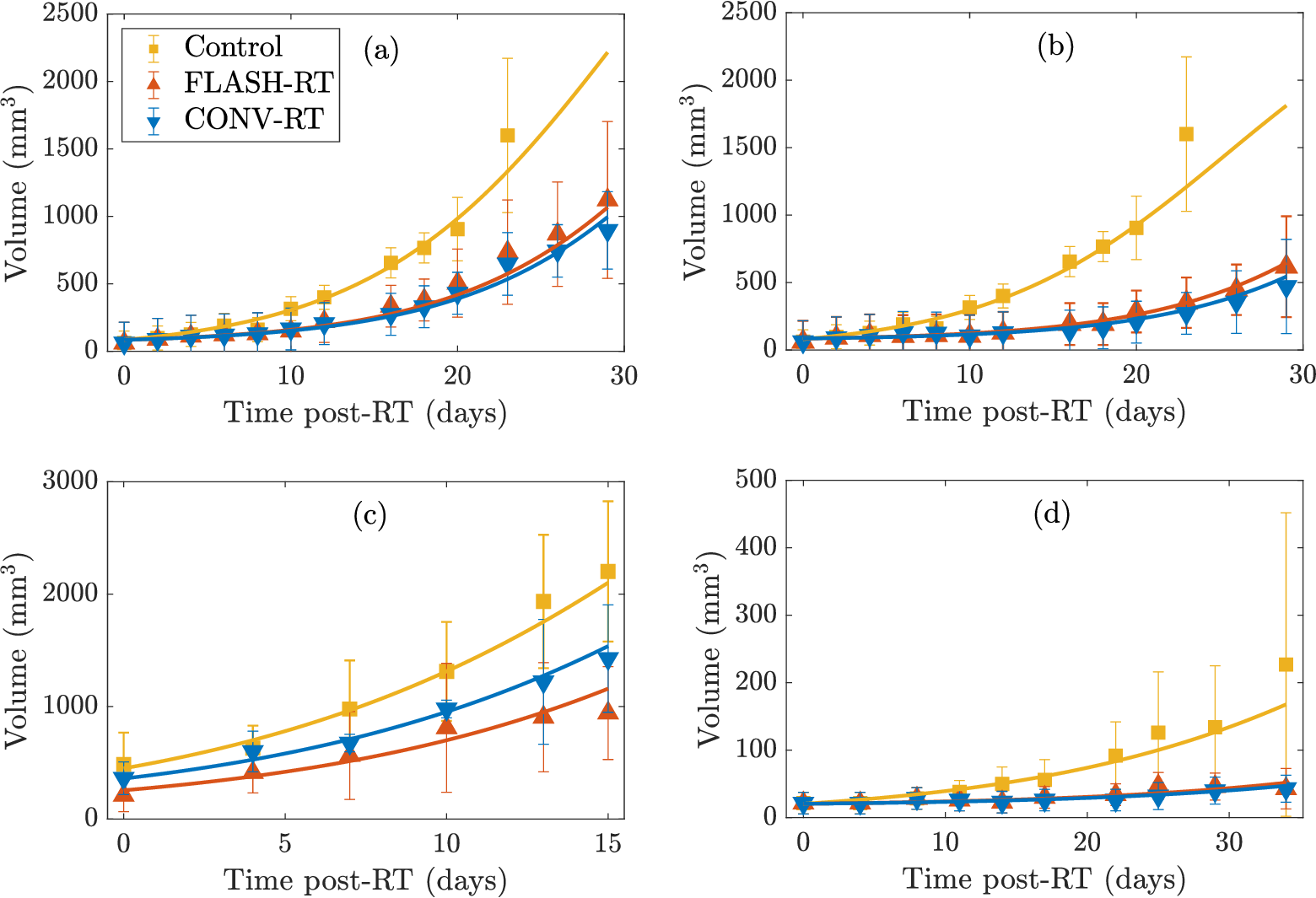}
	\caption{Best-fits of tumor growth curves (mean values and standard deviations) presented by Diffenderfer~\textit{et al.}~\cite{diffenderfer2020} (a, b) and Zhu~\textit{et al.}~\cite{zhu2023a} (c, d).}
	\label{fig_3}
\end{figure}

For the fit of Diffenderfer \textit{et al.} data, who irradiated the same tumor model with two different radiation doses, we used the same set of parameters for both datasets. Additionally, we used three distinct initial volumes, one for each curve of control, CONV-RT, and FLASH-RT, to emulate the noticeable differences in pre-irradiation volumes observed in one of the experiments reported in Zhu~\textit{et al}. The best-fitting curves and experimental data are shown in Figure~\ref{fig_3}. The best-fitting parameters for each experimental study are summarized in Table~\ref{table_s1}. 

\subsection{Analysis of CONV-RT and FLASH-RT iso-effectiveness from dose-volume curves}
\label{section_results_statistics}


We used the results of the previous section to investigate the significance of the difference between tumor growth curves. This study was performed as follows: i) a random growth curve was generated by sampling parameters from a normal distribution with the mean given by the best-fitting parameters reported in Table \ref{table_s1} and a 20~\% relative standard deviation, sampling the oxygenation from the $\Omega$ sample previously discussed, and assigning a treatment group (control, FLASH-RT, CONV-RT); ii) step i was repeated to obtain the sample size as used in each experimental study; iii) the ANCOVA methodology was used to detect significant differences between groups, as described in Section~\ref{section_materials_data}; iv) steps i--iii were repeated 1000 times to achieve enough statistics. This is illustrated in Figure \ref{fig_s4} for one of the thousand simulations.

A very large percentage of the simulated experiments showed a significant difference between control and FLASH-RT/CONV-RT. However, only a small fraction of the simulated experiments (16\% to 29\%) showed a significant difference between FLASH-RT/CONV-RT (see Figure \ref{fig_s5}). This happened in spite of the different surviving fractions associated with FLASH-RT and CONV-RT due to the ROD effect: the differences in surviving fractions were not large enough to cause significant differences in growth curves with sample sizes ranging from 5 to 15 animals.

\subsection{Estimation of the effects of ROD on tumor control probability}
\label{section_results_clinical}

\subsubsection{Homogeneously oxygenated tumors}

\begin{figure}[htb]
	\centering
	\includegraphics[width=\columnwidth]{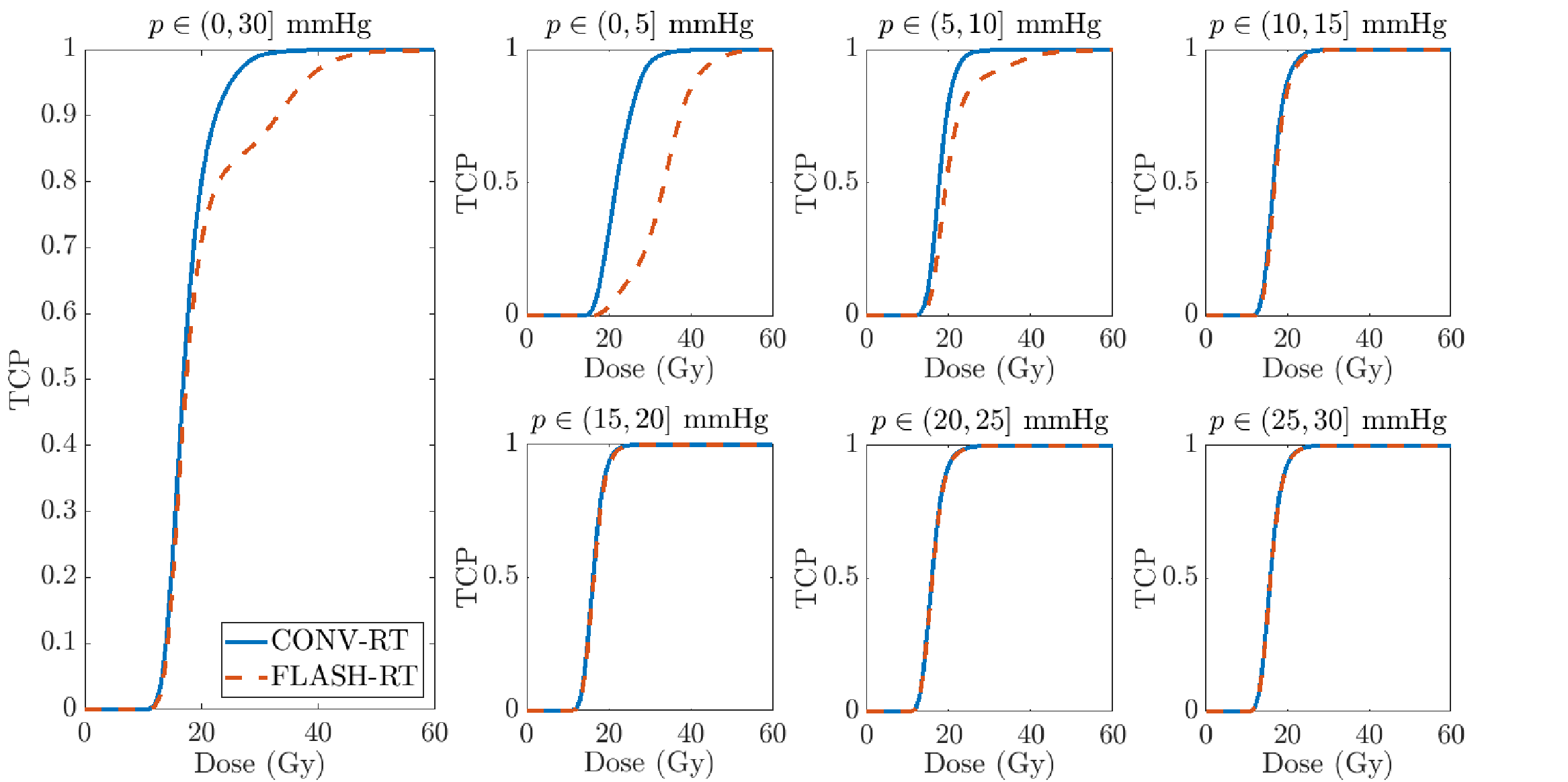}
	\caption{Simulated TCP-dose curves for homogeneously oxygenated tumors, with oxygenation $p$ ranging from 1 to 30 mmHg. The results are presented for the whole range, as well as in 6 groups with oxygenations ranging from (0, 5], (5, 10], (10, 15], (15, 20], (20, 25], and (25, 30] mmHg.}
	\label{fig_TCP_d}
\end{figure}

Firstly, we restricted our analysis to homogeneously oxygenated tumors in order to study the effect of the oxygenation level on CONV-RT and FLASH-RT. We obtained TCP-dose curves for CONV-RT and FLASH-RT treatments for the population of 1000 simulated tumors created in Section~\ref{section_materials_TCP}, but with homogeneous oxygenation ranging from 1 to 30~mmHg in steps of 1~mmHg. For each oxygen level, we solved the non-spatial form of Equation~(\ref{eq_oxyFlash}) without the diffusive term and obtained the respective TCP-dose curve for doses ranging from 2~Gy to 60~Gy in steps of 2~Gy. We used the parameters reported in Table \ref{table_0}, with $\alpha_\mathrm{ox}$=0.4~Gy$^{-1}$, $N_0$=10$^6$~cells (mean values, for each tumor the specific values were sampled from a normal distribution as discussed in Section~\ref{section_materials_TCP}), and $\alpha_\mathrm{ox}/\beta_\mathrm{ox}$=10~Gy. In Figure~\ref{fig_TCP_d} we present the TCP-dose curves for CONV-RT and FLASH-RT, and in Table~\ref{table_s2} we report the $D_{50}$ and $D_{90}$ values for each curve. The same trend observed in the study of the SF was found, with higher differences between FLASH-RT and CONV-RT appearing on hypoxic tumors.

\subsubsection{Heterogeneously oxygenated tumors}

We also performed the same study in more realistic tumors with heterogeneous oxygen distributions. The populations of heterogeneous tumors were created as discussed in Section \ref{section_materials_TCP}. Different $\alpha_\mathrm{ox}/\beta_\mathrm{ox}$ values were investigated: low ($\alpha_\mathrm{ox}/\beta_\mathrm{ox}$=3 Gy, $\alpha_\mathrm{ox}$=0.175 Gy$^{-1}$), medium ($\alpha_\mathrm{ox}/\beta_\mathrm{ox}$=10 Gy, $\alpha_\mathrm{ox}$=0.4 Gy$^{-1}$) and high ($\alpha_\mathrm{ox}/\beta_\mathrm{ox}$=20 Gy, $\alpha_\mathrm{ox}$=0.553 Gy$^{-1}$). The radiosensitivity of the cells was set to yield the same $D_{50}$ for CONV-RT in each case. TCP-dose curves are presented in Figure \ref{fig_tcp_a_b}. Curves for FLASH-RT show higher $D_{50}$ values, as reported in Table \ref{table_7}. When accounting for the effect of the different $\alpha_\mathrm{ox}/\beta_\mathrm{ox}$ by calculating the biologically effective dose (BED)~\cite{jones2001} associated to differences in $D_{50}$, the effective difference between FLASH-RT and CONV-RT grows with decreasing $\alpha_\mathrm{ox}/\beta_\mathrm{ox}$.

\begin{table}[htb]
	\centering
		\begin{tabular}{ccccc}
			\cline{1-5}
			$\alpha_\mathrm{ox}/\beta_\mathrm{ox}$ & $D_{50}^\textrm{C}$    & $D_{50}^\textrm{F}$ & $\Delta D_{50}$ & $\textit{BED}(\Delta D_{50})$\\ \hline
			\multicolumn{1}{c}{3}    & 29.17    & 31.52             & 2.35          & 4.19 \\
			\multicolumn{1}{c}{10}   & 29.17    & 31.78             & 2.61          & 3.29\\
			\multicolumn{1}{c}{20}   & 29.16    & 31.98             & 2.82          & 3.22 \\ \hline
		\end{tabular}
	\caption{Differences in D$_{50}$ between CONV-RT and FLASH-RT according to the $\alpha_\mathrm{ox}/\beta_\mathrm{ox}$ ratio. Differences are reported in grays ($\Delta D_{50}$) and as BED to account for the different $\alpha/\beta$.}
	\label{table_7}
\end{table}

\begin{figure}[htb]
	\centering
	\includegraphics[width=\columnwidth]{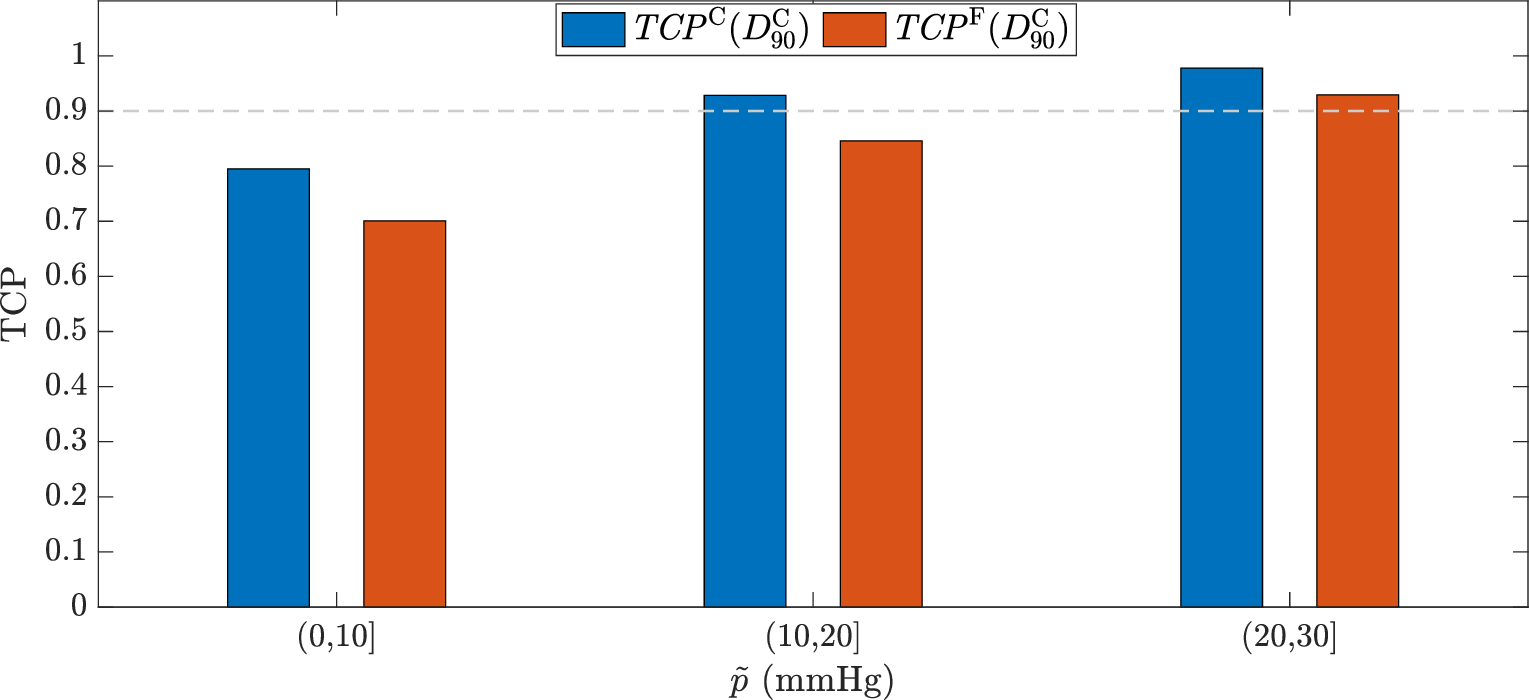}
	\caption{TCP values for CONV-RT ($\textit{TCP}^\text{C}$) and FLASH-RT ($\textit{TCP}^\text{F}$) and differences between them ($\Delta\textit{TCP}$) for a dose of 37.22 Gy (equivalent to $D_{90}$ for the whole population with CONV-RT) according to tumor oxygenation, which was characterized by the median of their oxygen distribution, $\tilde{p}$: \textit{poorly oxygenated} ($\tilde{p}\le$ 10 mmHg), \textit{moderately oxygenated} (10 $<\tilde{p}\le$ 20 mmHg), and \textit{well oxygenated} tumors ($\tilde{p} >$ 20 mmHg).}
	\label{fig_TCPf}
\end{figure}

In order to investigate the influence of oxygenation on the TCP, we split the tumors into three groups according to the median of their oxygen distribution, $\tilde{p}$. Namely, \textit{poorly oxygenated} ($\tilde{p}\le$ 10 mmHg), \textit{moderately oxygenated} (10 $<\tilde{p}\le$ 20 mmHg), and \textit{well oxygenated} tumors ($\tilde{p} >$ 20 mmHg). We then evaluated the effect of the $D_{90}$ obtained for the whole population (CONV-RT) on each individual group. This is shown in Figure \ref{fig_TCPf} (for $\alpha_\mathrm{ox}/\beta_\mathrm{ox}$=10 Gy) and Table \ref{table_TCP_10_v2}, where it can be noticed that tumors are controlled at different rates according to their oxygenation status, and that the loss of TCP caused by ROD in FLASH-RT is more noticeable in \textit{poorly oxygenated} tumors.

\subsubsection{Heterogeneously oxygenated tumors: effect of the oxygenation}

\begin{figure}[H]
	\centering
	\includegraphics[width=\textwidth]{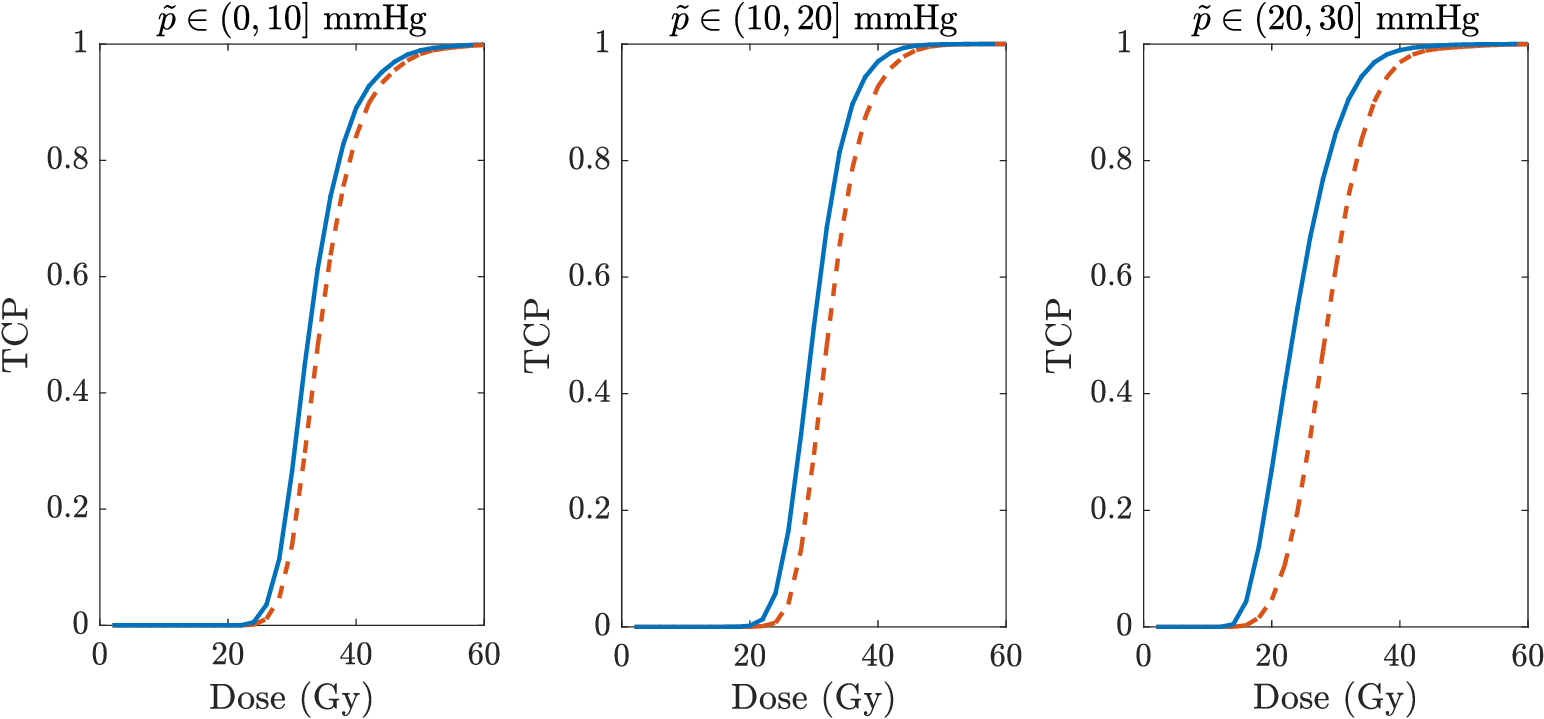}
	\caption{TCP-dose curves for CONV-RT (solid lines) and FLASH-RT (dashed lines) in tumors with heterogeneous oxygen levels according to their median oxygen partial pressure, $\tilde{p}$: \textit{poorly oxygenated} ($\tilde{p}\le$ 10~mmHg), \textit{moderately oxygenated} (10 $<\tilde{p}\le$ 20~mmHg), and \textit{well oxygenated} tumors ($\tilde{p} >$ 20~mmHg). The $\alpha_\mathrm{ox}/\beta_\mathrm{ox}$ ratio was 10~Gy.}
    \label{fig_TCPh}
\end{figure}

A final simulation experiment was performed. The TCP-dose curves of the whole population of heterogeneous tumors were calculated separately for each oxygenation level (as defined above), both for FLASH-RT and CONV-RT. The results are presented in Figure \ref{fig_TCPh}, and the quantification of differences in $D_{50}$ and $D_{90}$ are summarized in Table \ref{table_TCPh} (for $\alpha_\mathrm{ox}/\beta_\mathrm{ox}$=10 Gy). When analyzing the TCP-dose curves according to the oxygenation of the tumors, the differences between FLASH-RT and CONV-RT seem to be more important for well oxygenated tumors than for poorly oxygenated tumors (as qualitatively shown in the figure and quantitatively shown in the table), which may seem to conflict the results obtained for homogeneous tumors.

This apparent contradiction can be explained by the complex interplay between the heterogeneity of the oxygenation, the OER effect and the ROD effect. \textit{Well oxygenated} tumors are controlled with much lower doses than \textit{moderately/poorly oxygenated} tumors due to the OER effect. Tumor control is mostly affected by low oxygenation regions, which are present even in \textit{well oxygenated} tumors in our population (as shown in the oxygen histograms reported in Figure \ref{fig_s1}), and the ROD effect on the radiosensitivity is also more important in low oxygenation regions. In \textit{well oxygenated} tumors, a smaller fraction of cells is affected by the ROD effect, but the lower doses needed to achieve control in these tumors make the overall FLASH effect larger than for \textit{moderately/poorly oxygenated} tumors. Intuitively, a small fraction of tumor cells is shifted from $p\sim$ 4 mmHg to $p\sim$ 1 mmHg in well oxygenated tumors, while a large fraction of cells undergoes that shift for 
\textit{poorly oxygenated} tumors. However, that small fraction of cells cannot be killed by the dose $D\sim$ 32 Gy needed to achieve 90\% control for those tumors (causing a large drop of TCP for FLASH-RT and thus requiring a much larger dose to compensate for the ROD effect), while the large fraction of cells undergoing that shift for \textit{poorly oxygenated} tumors can still be adequately controlled with the dose $D\sim$ 40 Gy needed to achieve 90\% for those tumors (therefore causing a more modest drop of TCP and requiring less of a boost to compensate for the ROD effect).

These results may be of interest in a hypothetical clinical trial where tumors were assigned to groups receiving different doses according to the oxygenation/hypoxic status. If FLASH-RT was used to treat those groups, the ROD effect might have a more detrimental effect on well oxygenated tumors (receiving lower doses) than on poorly oxygenated tumors (receiving higher doses).

\begin{table}[h]
	\centering
		\begin{tabular}{ccccccc}
			\hline
			$\tilde{p}$ (mmHg)& $D_{50}^\textrm{C}$ (Gy) & $D_{50}^\textrm{F}$ (Gy) & $D_{50}^\textrm{F}/D_{50}^\textrm{C}$ & $D_{90}^\textrm{C}$ (Gy) & $D_{90}^\textrm{F}$ (Gy) & $D_{90}^\textrm{F}/D_{90}^\textrm{C}$\\ \hline
			$(0,10]$        & 32.56    & 34.23       & 1.05  & 40.43    & 42.06       & 1.04  \\
			$(10,20]$      & 29.83    & 32.19       & 1.08  & 36.11    & 38.86       & 1.08  \\
			$(20,30]$      & 23.30    & 28.36       & 1.22  & 31.79    & 35.96       & 1.13  \\ \hline
		\end{tabular}
	\caption{Differences in D$_{50}$ and D$_{90}$ between CONV-RT and FLASH-RT in tumors with heterogeneous oxygen levels according to their median oxygen partial pressures, $\tilde{p}$. The $\alpha_\mathrm{ox}/\beta_\mathrm{ox}$ ratio was 10 Gy.}
	\label{table_TCPh}
\end{table}

\section{Conclusions}
\label{section_conclusions}

In recent years, there has been significant interest in FLASH radiotherapy because \textit{in vivo} experiments have shown its potential to spare normal tissue while seemingly maintaining the effectiveness of conventional radiotherapy on tumors. Several studies have suggested that the FLASH effect arises from the radiolytic oxygen depletion process caused by the ultra-high dose rates of FLASH irradiation, which leads to increased cell radioresistance due to the oxygen enhancement effect. However, other studies claim that oxygen depletion alone is insufficient to fully explain the sparing effect. While it is not entirely clear what the mechanisms behind the FLASH effect are, it is known that FLASH-RT leads to ROD, both from experiments with different oxygen solutions and \textit{in vivo} studies~\cite{cao2021, el2022, ha2022, jansen2022}. Understanding why the protective effect of ROD is not observed in tumors remains an ongoing area of research, with potentially important clinical implications: as FLASH-RT moves closer to clinical application~\cite{bourhis2019a, bourhis2019b, vozenin2022}, it is important to guarantee that clinical trials of FLASH-RT will not lead to poor tumor control probabilities, and, if necessary, to boost the radiation dose to compensate for such effects.   

In this regard, priority has to go to a recent theoretical study by Liew \textit{et al.}~\cite{liew2023} that argued that FLASH-RT might in fact not preserve tumor control probability. In this independent study we further investigated this possibility by using mathematical modeling based on the differential equations governing the oxygen distribution in tissues, the linear-quadratic model with oxygen enhancement ratios to account for the oxygen effect on irradiated cells, a simple model of tumor growth after irradiation, and the standard LQ-Poisson TCP formulation to calculate the tumor control probability of different treatments. With these models, we systematically and progressively analyzed experimental data of ROD in solutions, ROD \textit{in vivo}, surviving fractions \textit{in vitro}, and tumor growth in preclinical irradiated tumors. We finally extrapolated the results obtained to the analysis of TCP-dose curves in a clinical setting, and the potential effect of ROD on the clinical iso-effectiveness of FLASH-RT (tumor control probabilities). 

Our study suggests that ROD may lead to differences in the surviving fraction of tumor cells between conventional and FLASH radiotherapies, which may not be large enough to induce significant differences in the volume evolution of preclinical tumors (that are far from control). Nonetheless, these differences may affect clinical TCPs, especially in low $\alpha/\beta$ tumors, where the ROD effect is expected to cause more cell sparing. In this regard, our study is in agreement with the results presented in~\cite{liew2023}.

Somewhat counterintuitive results were found when analyzing the effect of tumor oxygenation on the ROD effect. The ROD effect is more important in poorly oxygenated cells, a trend that our simulations reproduced for homogeneously oxygenated cells/tumors. Also, in a population of heterogeneously oxygenated tumors all irradiated with the same dose, the hypoxic tumors are those most affected by the loss of tumor control associated with the ROD effect. However, when analyzing the TCP-dose curves of those tumors according to their oxygenation status, we found that \textit{well oxygenated} tumors, which require a lower dose to be controlled, are more affected by the detrimental ROD effect than tumors with \textit{moderate} and \textit{poor} oxygenations (defined in this work as tumors with median oxygen partial pressures above and below 10~mmHg, respectively). These results may be of interest in a hypothetical clinical trial where tumors were prescribed a different dose according to the oxygenation/hypoxic status.

Certainly, it cannot be discarded that other effects that have not been modeled in this work, like radiation-induced immune effects or differences in the production yield of reactive species production between tumor and non-tumor cells~\cite{spitz2019, jin2020, abolfath2022, shukla23}, can contribute to tumor control and maintain the iso-effectiveness of FLASH radiotherapy. Ideally, it would be of special interest to experimentally investigate not only the volume-time curves of tumor irradiated with FLASH-RT/CONV-RT, but also the tumor control probabilities achieved with those treatments, yet this may not be achievable due to the low radiosensitivity of tumor models and the large number of animals that would be necessary. Regarding the transfer of FLASH-RT to the clinic, it might be appropriate to target first well oxygenated tumors with moderate and large $\alpha/\beta$ ratios.

\section*{Acknowledgements}

This work has received funding from Xunta de Galicia-GAIN (IN607D 2022/02) and Ministerio de Ciencia e Innovación-AEI (PLEC2022-009476). OLP acknowledges support from Ministerio de Ciencia e Innovación, project PID2021-122625OB-I00 with funds from MCIN/AEI/10.13039/501100011033/ ERDF, UE, and from Xunta de Galicia (2021 GRC Gl-1563 - ED431C 2021/15).

\renewcommand{\thefigure}{S\arabic{figure}}
\renewcommand{\thetable}{S\arabic{table}}
\setcounter{figure}{0}
\setcounter{table}{0}

\newpage

\appendix

\section{Supplementary tables}

\begin{table*}[h]
	\centering
		\begin{tabular}{cccc}
			\cline{1-4}
														& {Diffenderfer (a \& b)}  			& {Zhu (c)} 				& {Zhu (d)} \\ \hline
\multicolumn{1}{l}{$\alpha_\mathrm{ox}$~(Gy$^{-1}$)}    & 0.083                             & 0.012                       & 0.163  \\ 
\multicolumn{1}{l}{$\lambda$~(days$^{-1}$)}			    & 0.134                             & 0.120                       & 0.068  \\ 
\multicolumn{1}{l}{$K$~(mm$^{3}$)}          			& 4.80$\times10^3$             		& 7.86$\times10^3$            & 0.88$\times10^3$ \\ 
\multicolumn{1}{l}{$\phi$~(days$^{-1}$)}    			& 2.94$\times10^{-29}$       		& 1.92$\times10^{-11}$        & 2.71$\times10^{-18}$ \\ 
\multirow{3}{*}{$V_0$~(mm$^{3}$)}                       & \multirow{3}{*}{84.87}            & 449.98 (control)     & \multirow{3}{*}{20.54} \\
			&  &  359.42 (CONV-RT) & \\
			&  &  255.00 (FLASH-RT) & \\
			\hline
		\end{tabular}
	\caption{Best-fitting model parameters for the tumor growth curves reported by Diffenderfer~\textit{et al.}~\cite{diffenderfer2020} and Zhu~\textit{et al.}~\cite{zhu2023a}.}
	\label{table_s1}
\end{table*}

\begin{table*}[h]
	\centering
		\begin{tabular}{ccccccc}
			\hline
			$p$ (mmHg)& $D_{50}^\textrm{C}$ (Gy) & $D_{50}^\textrm{F}$ (Gy) & $D_{50}^\textrm{F}/D_{50}^\textrm{C}$ & $D_{90}^\textrm{C}$ (Gy) & $D_{90}^\textrm{F}$ (Gy) & $D_{90}^\textrm{F}/D_{90}^\textrm{C}$\\ \hline
			$(0,5]$        & 21.95    & 33.51       & 1.53  & 28.41    & 41.81    & 1.47  \\
			$(5,10]$       & 17.81    & 19.42       & 1.09  & 21.49    & 29.05    & 1.35  \\
			$(10,15]$      & 16.67    & 17.09       & 1.02  & 20.05    & 20.78    & 1.04  \\
			$(15,20]$      & 16.14    & 16.33       & 1.01  & 19.42    & 19.71    & 1.02  \\
			$(20,25]$      & 15.83    & 15.94       & 1.01  & 19.03    & 19.18    & 1.01  \\
			$(25,30]$      & 15.63    & 15.70       & 1.00  & 18.79    & 18.89    & 1.01  \\ \hline
		\end{tabular}
	\caption{Differences in D$_{50}$ and D$_{90}$ between CONV-RT and FLASH-RT in tumors with homogeneous oxygen distributions ranging from 1 to 30~mmHg. For simplicity, the oxygenations were merged in six groups.}
	\label{table_s2}
\end{table*}

\begin{table*}[htb]
	\centering
		\begin{tabular}{cccc}
		\hline
		    \multicolumn{4}{c}{$\alpha_\mathrm{ox}/\beta_\mathrm{ox}=3$ Gy}\\
			\hline
			$\tilde{p}$ (mmHg)& $\textit{TCP}^\text{C}(D_{90}^\textrm{C})$ & $\textit{TCP}^\text{F}(D_{90}^\textrm{C})$ & $\Delta\textit{TCP}$\\ \hline
			$(0,10]$         & 0.792    & 0.695       & 0.097  \\
			$(10,20]$       & 0.928    & 0.841       & 0.087  \\
			$(20,30]$       &   0.978    & 0.929       & 0.049  \\
		    \textit{all tumors}         & 0.900   & 0.828       & 0.072  \\
		    \hline
		    \hline
		    \multicolumn{4}{c}{$\alpha_\mathrm{ox}/\beta_\mathrm{ox}=10$ Gy} \\
			\hline
			$\tilde{p}$ (mmHg)& $\textit{TCP}^\text{C}(D_{90}^\textrm{C})$ & $\textit{TCP}^\text{F}(D_{90}^\textrm{C})$ & $\Delta\textit{TCP}$\\ \hline
			$(0,10]$         & 0.795    & 0.701       & 0.094  \\
			$(10,20]$       & 0.929    & 0.846       & 0.083  \\
			$(20,30]$       &   0.978    & 0.930       & 0.048  \\
		    \textit{all tumors}         & 0.900   & 0.831       & 0.069  \\
		    \hline
		    \hline
		    \multicolumn{4}{c}{$\alpha_\mathrm{ox}/\beta_\mathrm{ox}=20$ Gy} \\
			\hline
			$\tilde{p}$ (mmHg)& $\textit{TCP}^\text{C}(D_{90}^\textrm{C})$ & $\textit{TCP}^\text{F}(D_{90}^\textrm{C})$ & $\Delta\textit{TCP}$\\ \hline
			$(0,10]$         & 0.792    & 0.699       & 0.093  \\
			$(10,20]$       & 0.927    & 0.843       & 0.084  \\
			$(20,30]$       &   0.977    & 0.927       & 0.050  \\
		    \textit{all tumors}         & 0.900   & 0.830       & 0.070  \\\hline
		\end{tabular}
	\caption{TCP values  for CONV-RT ($\textit{TCP}^\text{C}$) and FLASH-RT ($\textit{TCP}^\text{F}$) and differences between them ($\Delta\textit{TCP}$) obtained for a dose equivalent to $D_{90}$ for the whole population with CONV-RT. Results are reported according to tumor oxygenation characterized by the median of their oxygen distribution, $\tilde{p}$: \textit{poorly oxygenated} ($\tilde{p}\le$ 10 mmHg), \textit{moderately oxygenated} (10 $<\tilde{p}\le$ 20 mmHg), and \textit{well oxygenated} tumors ($\tilde{p} >$ 20 mmHg).}
	\label{table_TCP_10_v2}
\end{table*}

\newpage

\section{Supplementary figures}

\begin{figure}[H]
	\centering
	\includegraphics[width=\columnwidth]{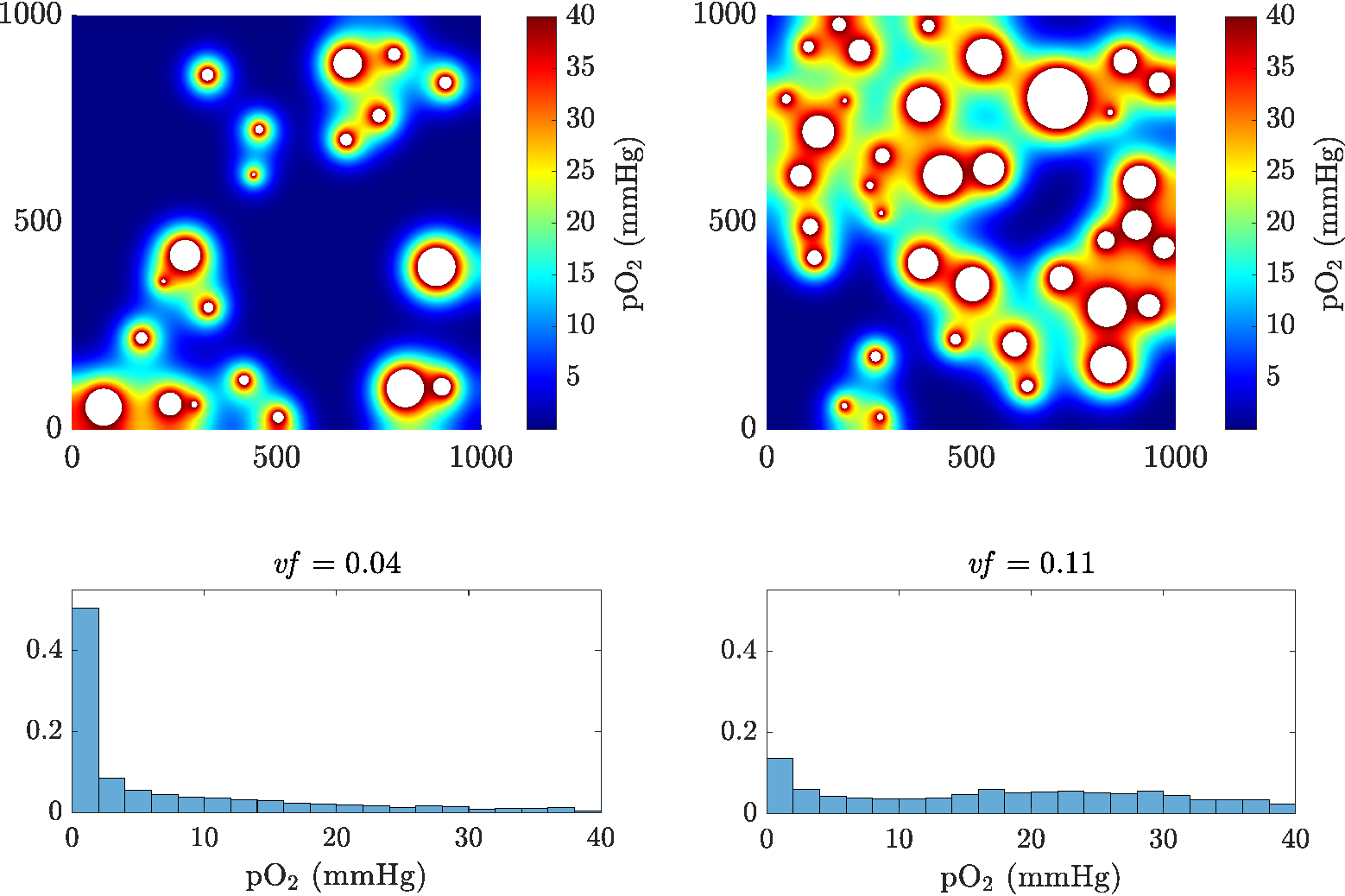}
	\caption{Example of solutions to Equation (\ref{eq_oxySteady}) on a squared domain of 1~mm$^2$~\cite{rodriguez2019} for different vascular fractions, leading to poorly and well oxygenated tumors. The top panels show the spatial distribution of the oxygenation and the distribution of capillaries, while the bottom panels show histograms of the oxygenation.}
	\label{fig_s1}
\end{figure}

\begin{figure}[H]
	\centering
	\includegraphics[width=\columnwidth]{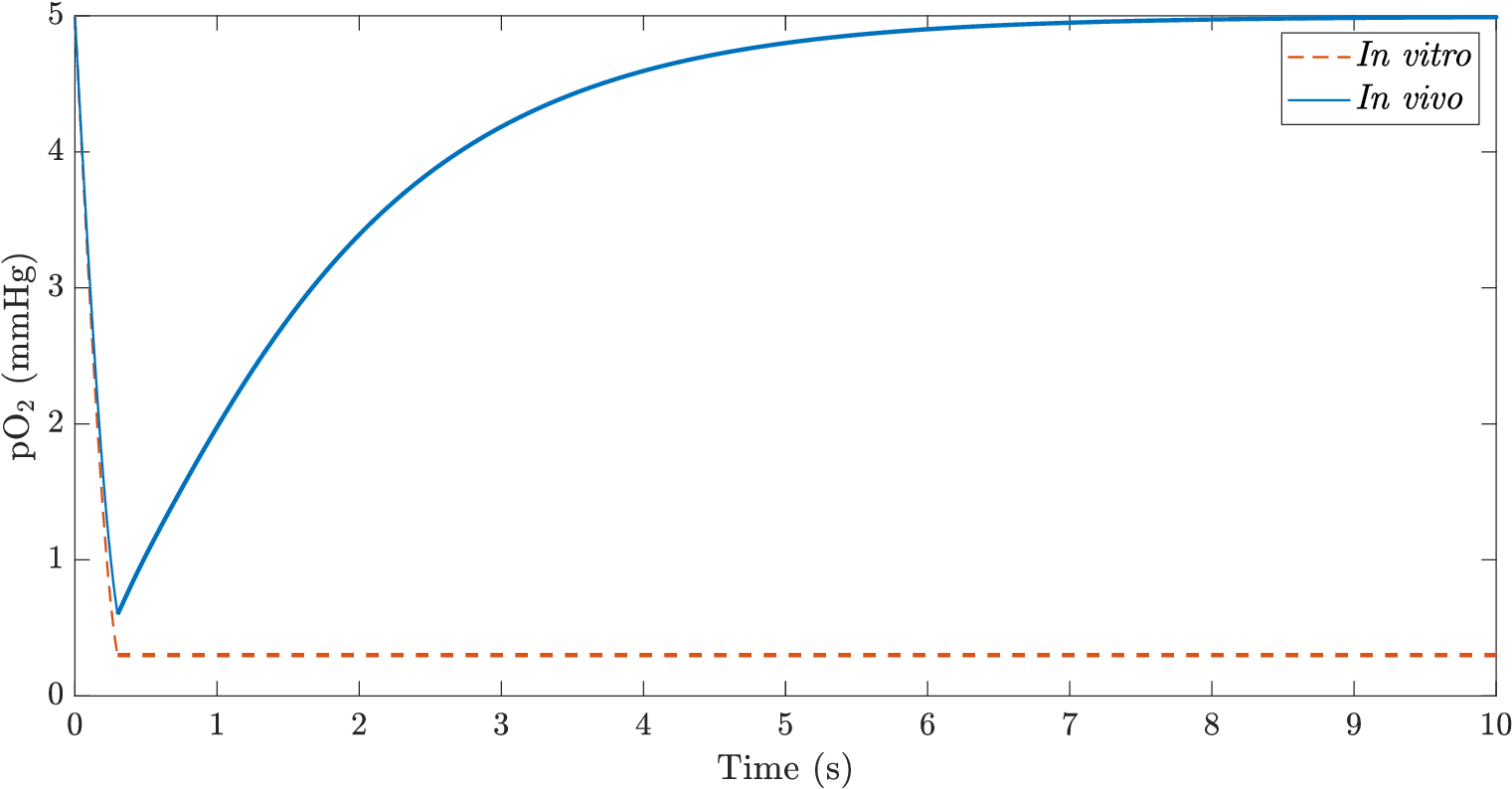}
	\caption{Example of solutions to Equation (\ref{eq_oxyFlash}) (oxygenation versus time) with an irradiation of 30 Gy (100 Gy s$^{-1}$) \textit{in vivo} (the capillaries act as a source and the initial oxygenation is recovered) and \textit{in vitro} (no recovery).}
	\label{fig_s2}
\end{figure}

\begin{figure}[H]
	\centering
	\includegraphics[width=\columnwidth]{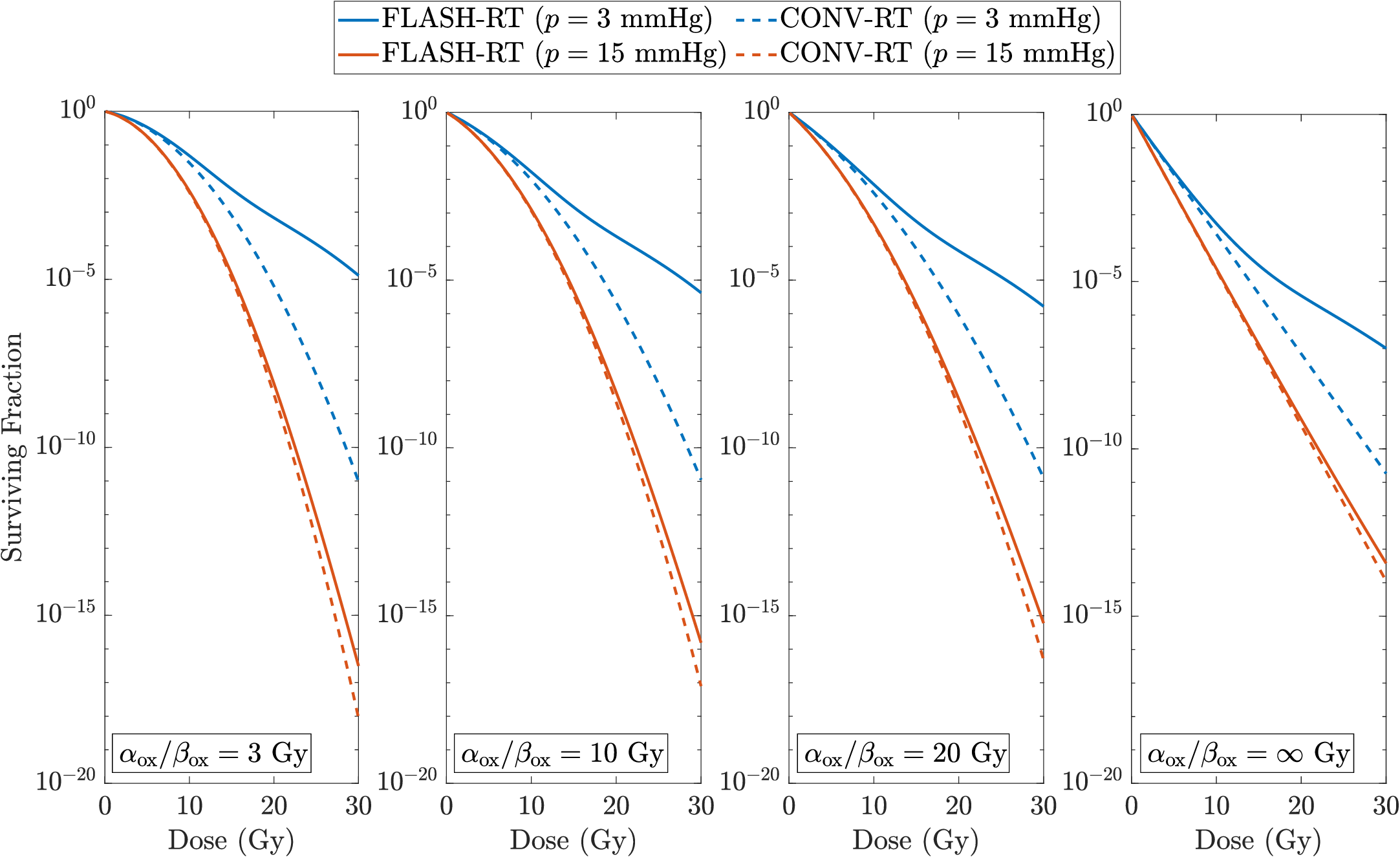}
	\caption{Surviving fraction versus dose curves for cells irradiated with CONV-RT and FLASH-RT at two oxygenation levels (poorly oxygenated, 3 mmHg, moderately well oxygenated, 15 mmHg). Results are presented for different values of $\alpha_\mathrm{ox}/\beta_\mathrm{ox}$, namely, 3, 10, 20 and $\infty$ Gy.}
	\label{fig_s3}
\end{figure}

\begin{figure}[H]
	\centering
	\includegraphics[width=\columnwidth]{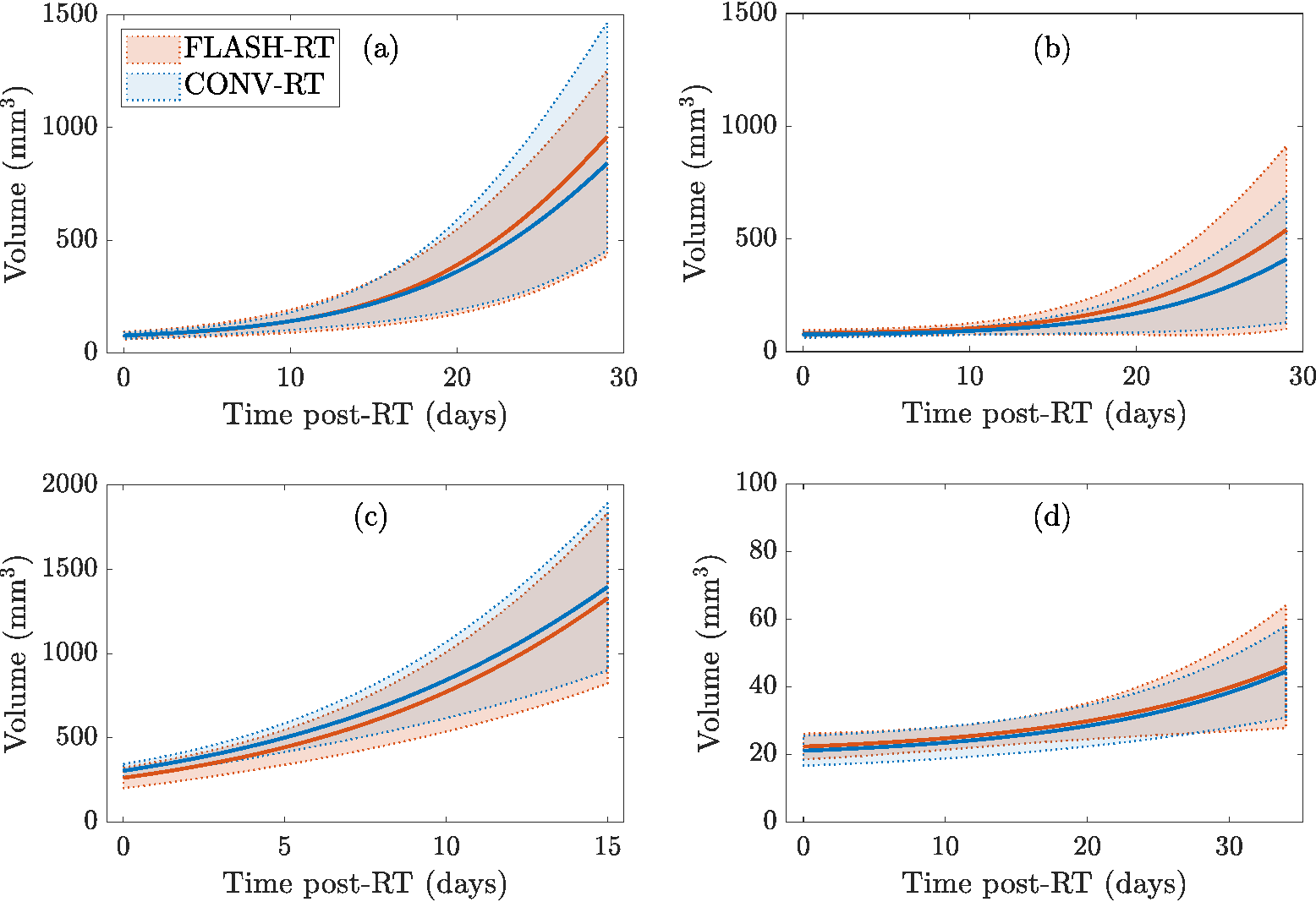}
	\caption{Example of the comparison between tumor growth curves after irradiation with FLASH-RT and CONV-RT from fits to the data reported by Diffenderfer~\textit{et al.}~\cite{diffenderfer2020} (a, b) and Zhu~\textit{et al.}~\cite{zhu2023a} (c, d). The solid lines represent the mean values and the shadow areas represent the standard deviation of 15 samples for FLASH-RT and CONV-RT (a), 14 samples for FLASH-RT and 15 for CONV-RT (b), 8 samples for FLASH-RT and CONV-RT (c and d).}
	\label{fig_s4}
\end{figure}

\begin{figure}[H]
	\centering
	\includegraphics[width=\columnwidth]{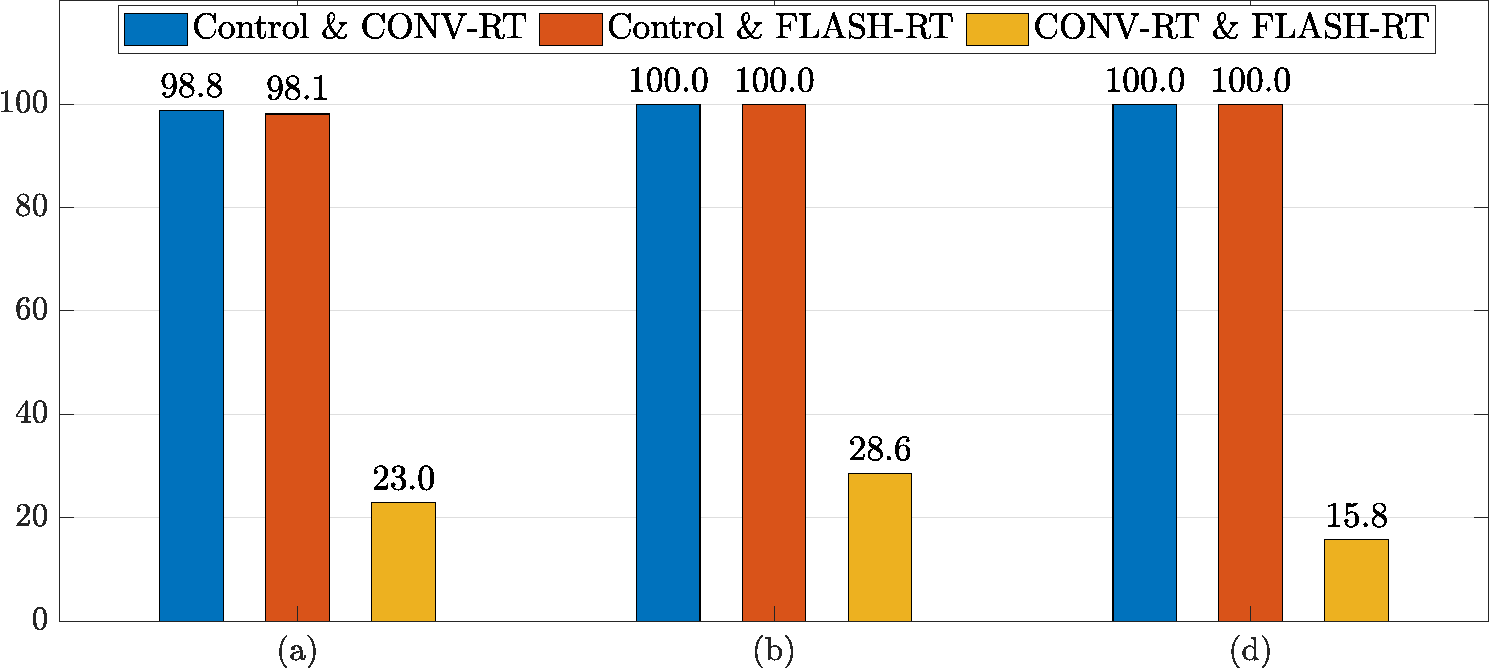}
	\caption{Percentage of simulated experiments showing a significant difference ($\textrm{p-value}<0.05$) between control, CONV-RT, and FLASH-RT groups, for simulations based on the experimental data reported by Diffenderfer~\textit{et al.}~\cite{diffenderfer2020} and Zhu~\textit{et al.}~\cite{zhu2023a}, and presented in Figure \ref{fig_3}(a), (b), and (d). The data from Zhu~\textit{et~al.} fitted in Figure \ref{fig_3}(c) were excluded from this analysis, because the baseline differences in volume between groups lead to significant differences that cannot be attributed to the treatments.}
	\label{fig_s5}
\end{figure}

\begin{figure}[htb]
	\centering
	\includegraphics[width=\columnwidth]{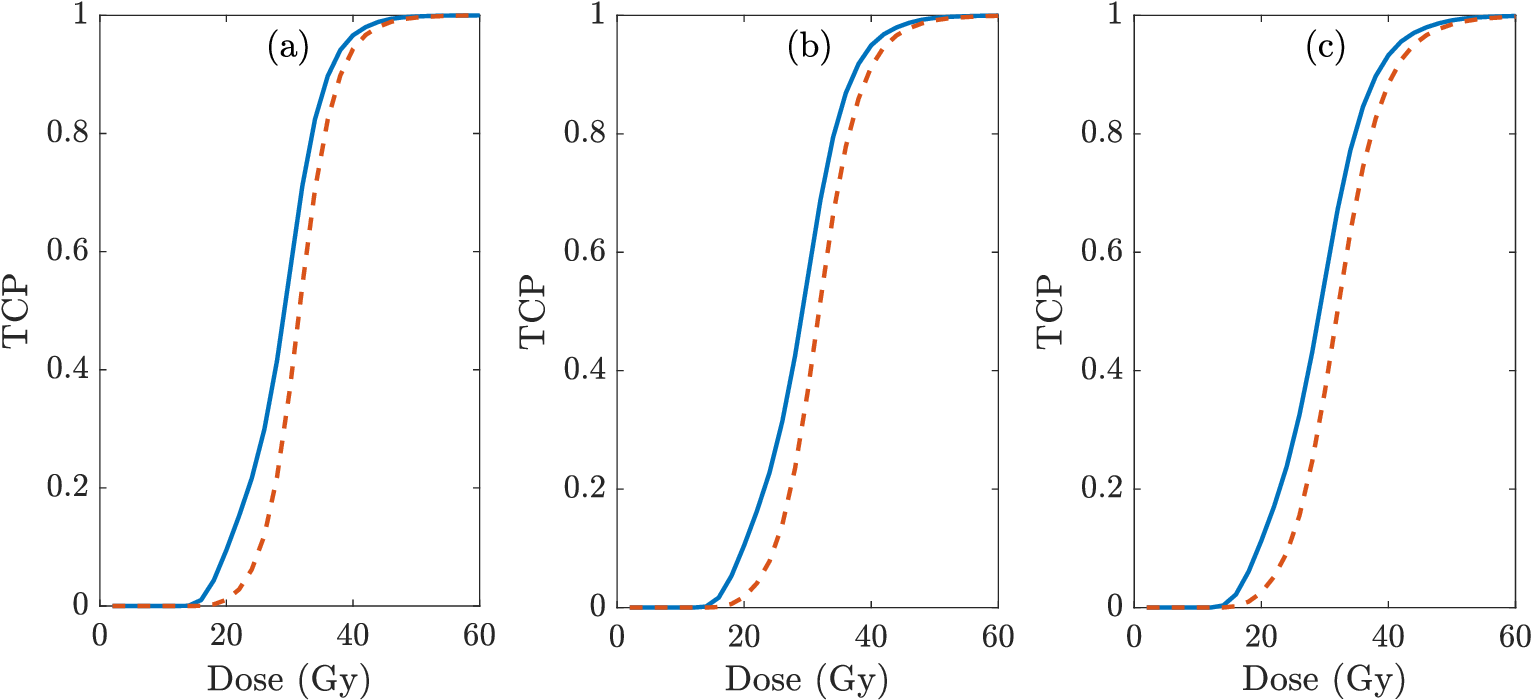}
	\caption{TCP-dose curves for heterogeneously oxygenated tumors with $\alpha_\mathrm{ox}/\beta_\mathrm{ox}$=3 Gy (a), 10 Gy (b) and 20 Gy (c) irradiated with CONV-RT (solid lines) and FLASH-RT (dashed lines).}
	\label{fig_tcp_a_b}
\end{figure}

\newpage

\bibliographystyle{unsrt}

\end{document}